\documentclass{article}
\usepackage[utf8]{inputenc}
\usepackage{spconf}
\usepackage{float}
\usepackage{graphicx}
\usepackage{multirow}
\usepackage{graphicx}
\usepackage{hyperref}
\usepackage{subcaption}
\usepackage{graphicx}
\usepackage{algorithm}
\usepackage{algpseudocode}
\usepackage{amsmath}
\DeclareRobustCommand{\IEEEauthorrefmark}[1]{\smash{\textsuperscript{\footnotesize #1}}}
\usepackage{enumitem}

\title{A New Approach to Extract Fetal Electrocardiogram Using Affine Combination of Adaptive Filters}
\name{%
\begin{tabular}{@{}c@{}}
Yu Xuan$^{\star}$\IEEEauthorrefmark{1} \qquad 
Xiangyu Zhang$^{\star}$\IEEEauthorrefmark{2}\qquad
Shuyue Stella Li$^{\dagger}$\IEEEauthorrefmark{2}\qquad \\
Zihan Shen$^{\dagger}$\IEEEauthorrefmark{3}\qquad
Xin Xie$^{\dagger}$\IEEEauthorrefmark{1}\qquad
Leibny Paola Garcia\IEEEauthorrefmark{2}\qquad
Roberto Togneri\IEEEauthorrefmark{4}
\end{tabular}}
\address{
\IEEEauthorrefmark{1}University of California San Diego, 
\IEEEauthorrefmark{2}Johns Hopkins University\\
\IEEEauthorrefmark{3}University of Chinese Academy of Sciences
\IEEEauthorrefmark{4}University of Western Australia
}

\begin{document}

\maketitle 
\begingroup
\def\thefootnote{$\star$ $\dagger$}
\footnotetext{Equal contribution in alphabetical order}
\endgroup

\begin{abstract}
The detection of abnormal fetal heartbeats during pregnancy is important for monitoring the health conditions of the fetus. While adult ECG has made several advances in modern medicine, noninvasive fetal electrocardiography (FECG) remains a great challenge. In this paper, we introduce a new method based on affine combinations of adaptive filters to extract FECG signals. The affine combination of multiple filters is able to precisely fit the reference signal, and thus obtain more accurate FECGs. We proposed a method to combine the Least Mean Square (LMS) and Recursive Least Squares (RLS) filters. Our approach found that the Combined Recursive Least Squares (CRLS) filter achieves the best performance among all proposed combinations. In addition, we found that CRLS is more advantageous in extracting FECG from abdominal electrocardiograms (AECG) with a small signal-to-noise ratio (SNR). Compared with the state-of-the-art MSF-ANC method, CRLS shows improved performance. The sensitivity, accuracy and F1 score are improved by 3.58\%, 2.39\% and 1.36\%, respectively.
\end{abstract}


\section{Introduction}
\begin{figure}
\centering
\includegraphics[scale = 0.46]{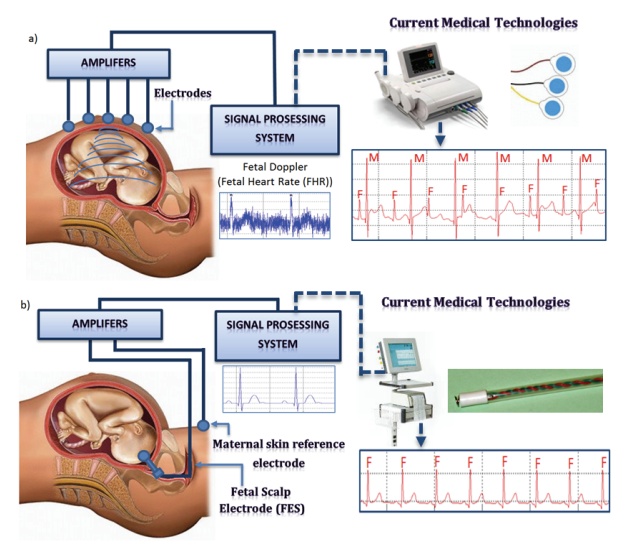}
\caption{FECG monitoring a) Non-invasive b) Invasive \cite{abdul2022review}}\vspace{-4mm}
\label{FECG}
\end{figure}

Fetal electrocardiography (FECG) is an important tool used by clinicians to track fetal cardiac status \cite{b} and can be effectively used to help clinicians make correct and responsive decisions during pregnancy or at delivery. Currently, there are two methods (Fig \ref{FECG}) to capture FECG: one is the \textbf{invasive scalp electrode} \cite{c} method, which directly measures the FECG signal, but can only detect the FECG signal during delivery. Besides, the scalp electrode is invasive and may cause serious harm to both the mother and the fetus. The other method uses a \textbf{non-invasive abdominal electrode} \cite{d}. An electrode is placed on the mother's abdomen to collect the abdominal body surface signal. However, the abdominal body surface signals of pregnant women are rather complex, including not only the weak FECG and maternal electrocardiography (MECG), but also the mother's respiratory noise and industrial frequency interference. In addition, the amplitude of MECG in the abdominal signal is often 2-10 times higher than that of FECG \cite{e}, making the extracted FECG extremely noisy. 

In this paper, we introduce an affine combination approach for ECG separation using multiple adaptive filters. Unlike previous combinations of multiple adaptive filters, the FECG signal extracted by the affine combination is closer to the true value and more complete. In addition, previous work mainly focuses on exploring the combination of only LMS filters with different step-size strategies. This is because these filters allow the combination to minimize their own quadratic errors by simply updating filter weights according to LMS rules. However, it is known that the performance of single Recursive Least Squares (RLS) is better than single LMS (and its family) \cite{yu2013performance} on most denoising tasks because RLS is more adaptable to non-stationary signals. The signal is generally non-stationary in most practical application scenarios, and the objective function of RLS is the exponentially weighted sum in the sense of least squares \cite{alexander1988analysis}. I.e., the objective function that the RLS algorithm minimizes is the weighted squared sum of error signals under the current coefficient for all signals, while LMS only considers the square of the instantaneous error signals. Therefore, RLS works better for non-stationary signals, including ECG signals. In this paper, we add RLS filters into a combination structure to extract FECG. To sum up, the contributions of this work include:

\begin{itemize}[noitemsep,topsep=0pt,leftmargin=10pt]
\setlength{\itemsep}{1pt}
\setlength{\parsep}{1pt}
\setlength{\parskip}{1pt}
\item A new FECG extraction method is proposed to obtain a more complete and accurate FECG signal by combining multiple filters. Our method improves the nonlinear fitting ability of the adaptive filter and has a good denoising ability.
\item We propose CRLS filters as an effective algorithm for FECG extraction, especially in low SNR scenarios. 
\item Compared with the state-of-the-art MSF-ANC method~\cite{n}, CRLS improves sensitivity, accuracy and F1 score by 3.58\%, 2.39\% and 1.36\%, respectively. We make our code available on GitHub to support future explorations\footnote{\url{https://github.com/scholesy123/ECG-signal}}.
\end{itemize}

\section{Related Work}\vspace{-1mm}
\subsection{FECG Extraction}\vspace{-1mm}

There are many methods and approaches proposed for the detection of FECG signals. And they can be roughly divided into two major categories - denoising and signal separation. 

First, treating the maternal signal as noise and then filtering out the maternal signal using methods such as matched filtering~\cite{f}, adaptive filtering~\cite{g}, and modified adaptive filtering~\cite{h}. RLS-based adaptive noise cancellation (ANC) method has also been used to remove MECG in non-invasive FECG extraction~\cite{u}. Combination-based approaches try to combine multiple LMS filters together, in which
the step size on the LMS filter can affect the performance of the combined LMS filter on FECG extraction. Thus, the combination strategy can be evaluated using qualitatively methods~\cite{w}\cite{x}. Moreover, the multiple sub-filter methods can be applied to extract FECG and for acoustic echo, cancellation \cite{pauline2022robust}~\cite{g}~\cite{t}. 

Second, the maternal signal and the fetal signal are treated as two independent signals and separated by certain methods such as blind source separation (BSS), independent component analysis (ICA), and singular value decomposition (SVD)~\cite{l}~\cite{j}. However, these methods usually impose strong, unrealistic assumptions that the signal and noises are mixed in a stationary manner \cite{sameni2008extraction}.

\vspace{-1mm}
\subsection{Convex Combination}\vspace{-1mm}
Another work~\cite{q} studies the mean square performance of the convex combination of two transversal LMS filters with different parameters to outperform both of them~\cite{y} and was applied to practical problems~\cite{z}. The affine combinations for adaptive filtering allow for different algorithms and even different adaptive filter lengths to be combined~\cite{aa}.

\section{Methodology}
\begin{figure}[h]
\centering
\includegraphics[scale = 0.65]{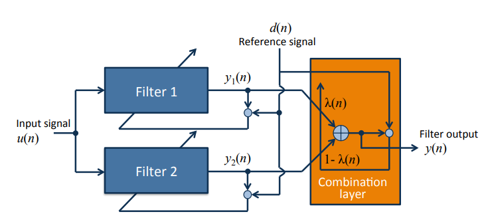}
\caption{combination filter scheme\cite{arenas2015combinations}}
\label{combine filter}\vspace{-2mm}
\end{figure}

\vspace{-2mm}
Fig \ref{combine filter} shows the general scheme of an affine combination of two adaptive filters. Filter 1 and Filter 2 can be either LMS or RLS. In total, three combinations are implemented and compared. \vspace{-2mm}

\subsection{Combined Least Mean Square (CLMS) Filter}\label{method:clms}
The first combination is CLMS, which is an affine combination of two LMS filters. One has a larger step size to achieve faster convergence and good tracking ability in situations of rapid change. The other has a smaller step size and performs well in stationary situations. The coefficients of two LMS filters are adapted to minimize their own quadratic errors using the standard LMS rule. The objective function $a(n)$ is updated at each iteration to minimize the quadratic error of the combined filter, also by means of a minimum square error stochastic gradient algorithm with step size $\mu_{a}$. The parameter $\eta(n)$ is defined via a sigmoid function. There are two benefits of using this sigmoid function: (1) it constrains the value of lambda between 0 and 1; (2) the $\eta(n)(1-\eta(n))$ term in its derivative reduces the adaptation speed and the gradient noise near the endpoints 0 and 1. The combined scheme is expected to perform similar to the fast and slow filters without degradation \cite{n}. \vspace{-4mm}

In the initialization step, the value of $a(n)$ is limited to the interval $[-4, 4]$, so the algorithm does not terminate when either $\eta_(n)$ or $1-\eta(n)$ are too close to 0. Furthermore, the parameter $\mu_{a}$ must be fixed to a very high value so that the combination is adapted even faster than the fastest LMS filter~\cite{o}. Note that, since $\eta(n)\in (0,1)$, the stability of the combined filter is guaranteed as long as the individual stability conditions of both filter 1 and filter 2 are satisfied.\vspace{-2mm}

\subsection{CRLS and RLS-LMS}\vspace{-1mm}
The second combination, CRLS, is an affine combination of the two RLS filters. One RLS filter has a larger forgetting factor to achieve faster adaption to situations of rapid changes. The other has a smaller forgetting factor and performs well in stationary situations~\cite{p}. \\
The third combination, RLS-LMS, is an affine combination of an RLS filter and an LMS filter. One filter is an RLS with faster convergence; the other is an LMS filter with better tracking ability. Note that RLS-LMS presents better tracking performance than CLMS~\cite{v}. The pseudo-code of this algorithm is shown below in Algorithm \ref{alg:code}.
All other parameter update mechanisms are consistent with Section \ref{method:clms}.

\begin{algorithm}
\caption{\label{alg:code}Combination of Adaptive Filters Algorithm}
Initialize $\mu_{1}$, $\mu_{2}$, $\mu_{a}$, $a^{+}$, $w_{1}(0)$, $w_{2}(0)$, $a(0)$

\textbf{Loop:n = 1 : end do}

\hspace{0.4cm} $\boldsymbol y_{i}(n)=\boldsymbol w_{i}(n)^{T}\boldsymbol u(n)$ \quad (i = 1, 2)

\hspace{0.4cm} $e_{i}(n)=d(n)-y_{i}(n)$

\hspace{0.4cm} $y(n)=\eta(n)y_{1}(n)+(1-\eta(n))y_{2}(n)$

\hspace{0.4cm} $e(n)=d(n)-y(n)$

\hspace{0.4cm} \textbf{If} CLMS:

\hspace{0.8cm} $w_{i}(n)=w_{i}(n-1)+\mu_{i}(n)e_{i}(n)\boldsymbol u(n)$

\hspace{0.4cm} \textbf{else if} CRLS:

\hspace{0.8cm} $\pi_{i}(n)=\frac{1}{\lambda_{i}}P(n-1)\boldsymbol u(n)$

\hspace{0.8cm} $k_{i}(n)=\frac{\pi_{i}(n)}{1+\boldsymbol u^{T}(n)\pi_{i}(n)}$

\hspace{0.8cm} $w_{i}(n)=w_{i}(n-1)+k_{i}(n)(d(n)-w_{i}^{T}(n-1)\boldsymbol u(n))$

\hspace{0.8cm} $P_{i}(n)=\frac{1}{\lambda_{i}}P_{i}(n-1)-\frac{1}{\lambda_{i}}k(n)\boldsymbol u^{T}(n)P_{i}(n-1)$

\hspace{0.4cm} \textbf{else if} RLS-LMS:

\hspace{0.8cm} $\pi(n)=\frac{1}{\lambda}P(n-1)\boldsymbol u(n)$

\hspace{0.8cm} $k(n)=\frac{\pi(n)}{1+\boldsymbol u^{T}(n)\pi(n)}$

\hspace{0.8cm} $w_{1}(n)=w_{1}(n-1)+k(n)(d(n)-w_{1}^{T}(n-1)\boldsymbol u(n))$

\hspace{0.8cm} $P(n)=\frac{1}{\lambda}P(n-1)-\frac{1}{\lambda}k(n)\boldsymbol u^{T}(n)P(n-1)$

\hspace{0.8cm} $w_{2}(n)=w_{2}(n-1)+\mu e_{2}(n)\boldsymbol u(n)$

\hspace{0.4cm} $e_{\alpha}(n)=e(n)(y_{1}(n)-y_{2}(n))$

\hspace{0.4cm} $a(n)=a(n-1)+\mu_{a}e_{\alpha}(n)\eta(n)(1-\eta(n))$

\hspace{0.4cm} $\eta(n)=\frac{1}{1+e^{-a(n)}}$

\hspace{0.4cm} \textbf{if} $a(n) < -a^{+}$

\hspace{0.8cm} $a(n)=-a^{+}$

\hspace{0.8cm} $\eta(n)=0$

\hspace{0.4cm} \textbf{else if} $a(n) > a^{+}$

\hspace{0.8cm} $a(n)=a^{+}$

\hspace{0.8cm} $\eta(n)=1$

\hspace{0.4cm} \textbf{end}

\hspace{0.4cm} $w(n)=\eta(n)w_{1}(n)+(1-\eta(n))w_{2}(n)$

\hspace{0.4cm} let $n=n+1$

\textbf{end}
\end{algorithm}\vspace{-5mm}
\section{Results and Discussion}\vspace{-3.5mm}

\subsection{Dataset}\vspace{-1.5mm}
We use a public dataset from DaIsy\footnote{\url{https://homes.esat.kuleuven.be/~smc/daisy/daisydata.html}} named ``Cutaneous potential recordings of a pregnant woman." There are eight single channels. The first five channels are the abdominal signals and the remaining three are the thoracic signals. The data size is $2500\times8$, the sampling time is 10s, and the sampling frequency is 250Hz. In this paper, we mainly use the second channel and eighth channel as input and reference signals.

\vspace{-2mm}\subsection{SNR Analysis}
We use the SNR to quantify the relative proportion of the effect from MECG to the effect of FECG signals. Generally, the larger the SNR is, the better the performance of the filter is. However, the opposite is true in this experiment because we are trying to obtain the weaker FECG, which is treated as noise in a typical SNR. Because the amplitude and power of the effect from the MECG as a noise signal is much larger than that of FECG, the SNR is calculated as
\begin{equation}
SNR=10 \log \left(\frac{P_{\text {signal }}}{P_{\text {noise }}}\right)=10 \log \left(\frac{P_{\text {FECG}}}{P_{\text {output }}}\right),
\end{equation}
where $P_{FECG}$ is the effect from FECG with any remaining, unfiltered MECG noise, and $P_{output}$ is the effect from MECG that's the output of the filter. Therefore, a small SNR indicates that the filter was able to filter out more of the MECG effect.
Because once the FECG is mixed with unsuppressed MECG, a large numerator is obtained, which is an undesired result. Meanwhile, the large value of the denominator indicates that the output of the filter is closer to the real MECG. Therefore, the smaller the SNR the better the noise suppression performance. SNR of different adaptive filters is shown in table \ref{table:1}.\vspace{-2mm}

\begin{table}[h!]
\centering
\begin{tabular}{lr} 
\hline 
Adaptive Filter&SNR(dB)\\
\hline  
NLMS&-4.1148\\
RLS&-11.1070\\
CLMS&-11.7637\\
RLS+LMS&-12.3267\\
CRLS&-12.4216\\
\hline
\end{tabular}
\caption{SNR of adaptive filters, CRLS has the best (lowest) SNR among all filter combinations.\vspace{-2mm}}
\label{table:1}
\end{table}
The CRLS has the smallest SNR value, which means that the CRLS has the best performance on noise suppression among all the filters. Also, the SNR of any combination of adaptive filters is lower than that of single filters.

\begin{figure}[htb]
     \centering
     \begin{subfigure}[b]{0.15\textwidth}
        \centering
        \includegraphics[width=\textwidth]{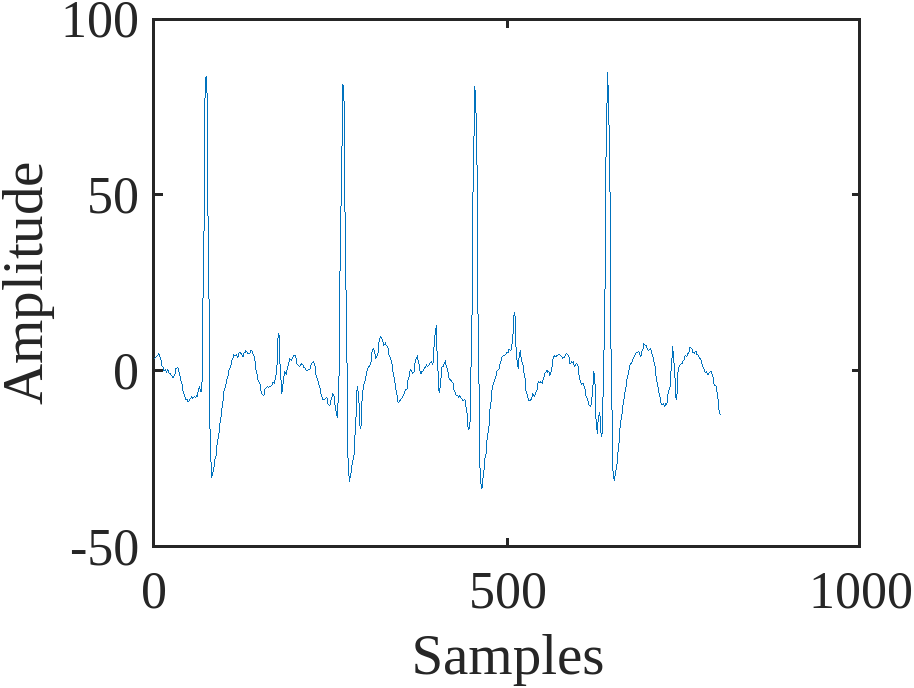}
        \caption{Input AECG}
        \label{fig:aecg}
    \end{subfigure}
    \hfil
    \begin{subfigure}[b]{0.15\textwidth}
         \centering
         \includegraphics[width=\textwidth]{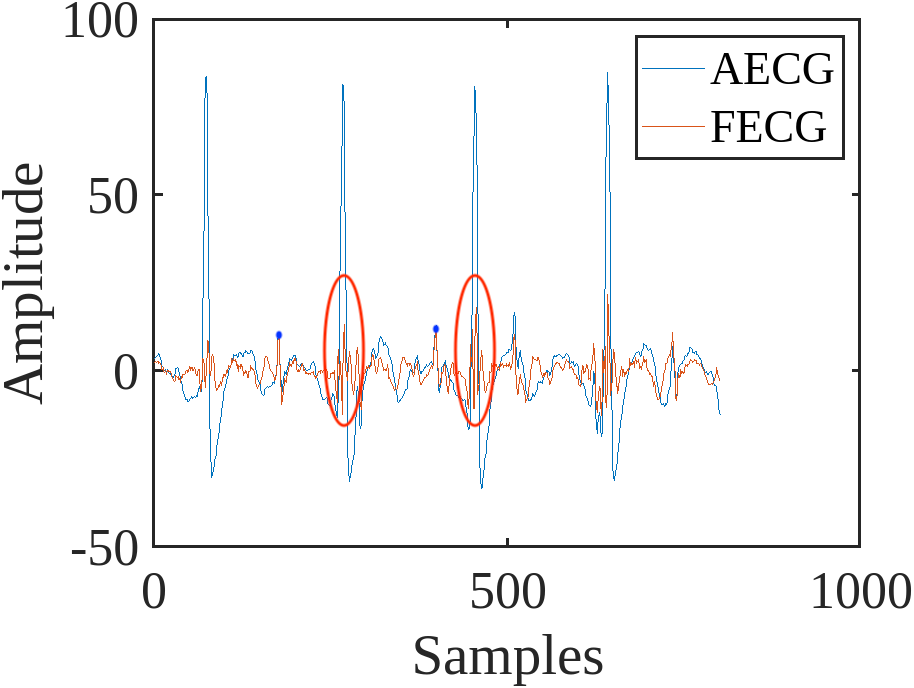}
         \caption{RLS result}
         \label{fig:rls}
    \end{subfigure}
    \hfil
    \begin{subfigure}[b]{0.15\textwidth}
         \centering
         \includegraphics[width=\textwidth]{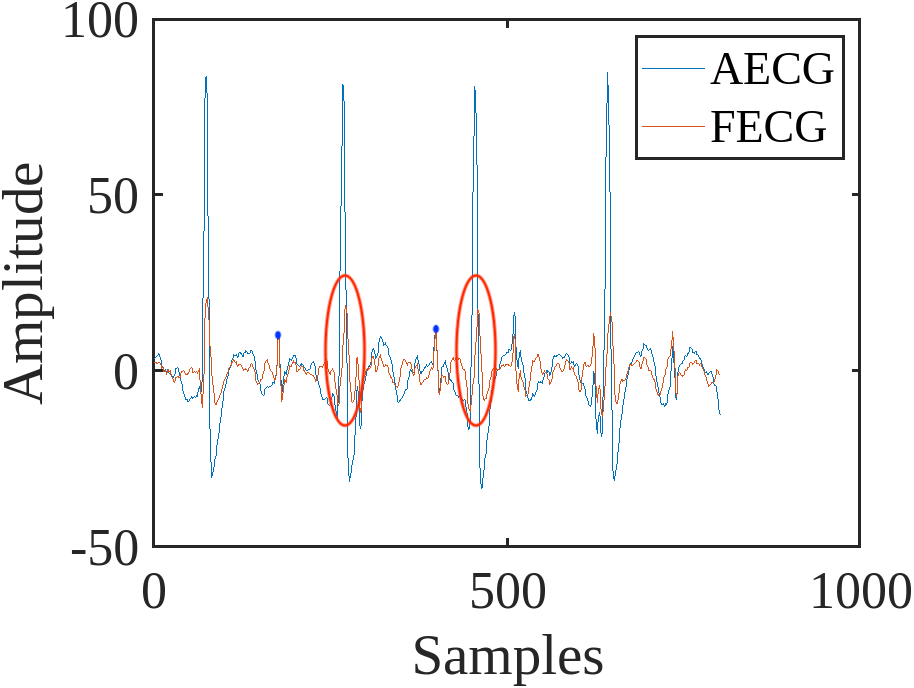}
         \caption{NLMS result}
         \label{fig:nlms}
    \end{subfigure}
    \hfil
     \begin{subfigure}[b]{0.15\textwidth}
         \centering
         \includegraphics[width=\textwidth]{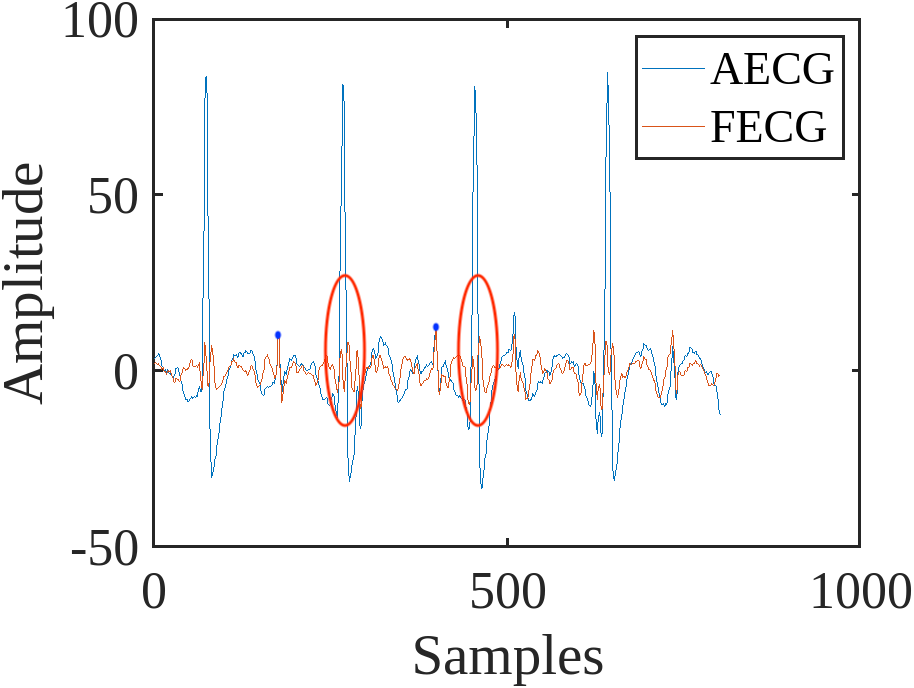}
         \caption{CLMS result}
         \label{fig:clms}
    \end{subfigure}
    \hfil
    \begin{subfigure}[b]{0.15\textwidth}
         \centering
         \includegraphics[width=\textwidth]{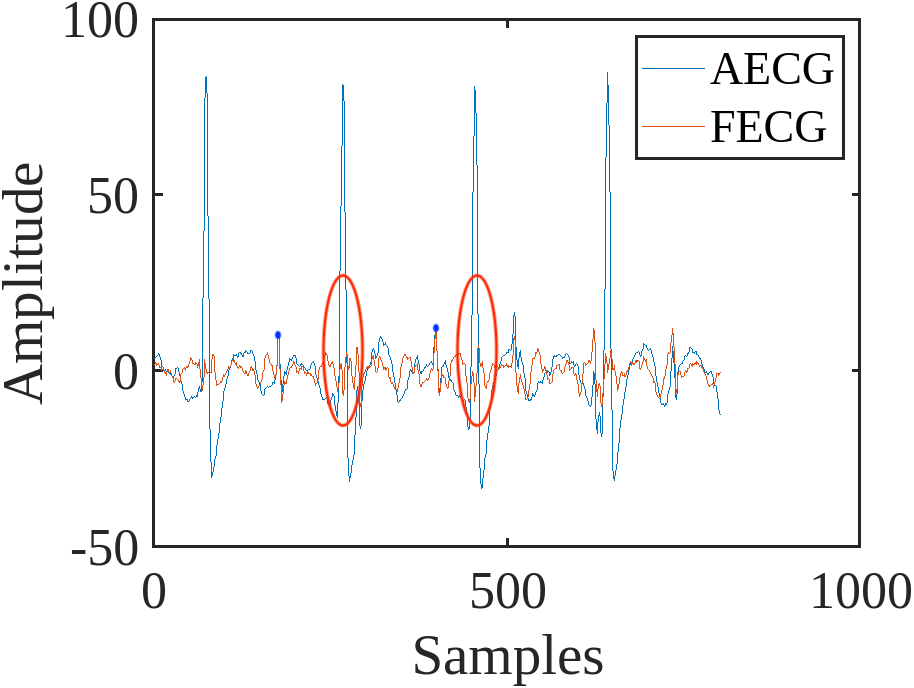}
         \caption{CRLS result}
         \label{fig:crls}
    \end{subfigure}
    \hfil
    \begin{subfigure}[b]{0.15\textwidth}
        \centering
        \includegraphics[width=\textwidth]{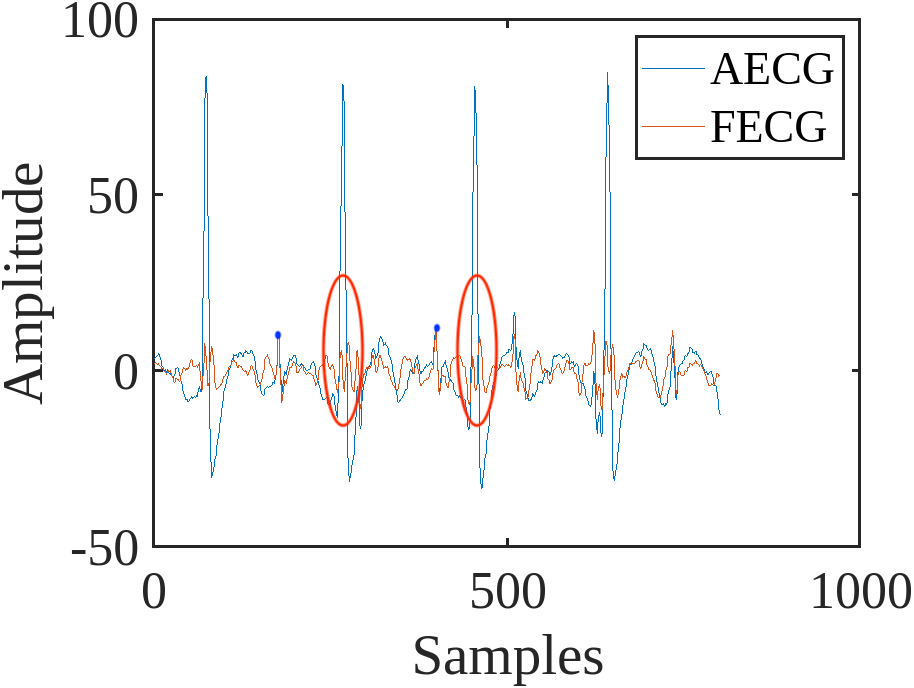}
        \caption{RLS-LMS result}
        \label{fig:rls_lms}
    \end{subfigure}\vspace{-1mm}
    \caption{Samples from 1600 to 2400 of AECG and FECG extracted by using adaptive filters\vspace{-2mm}}
    \label{six_graph}
\end{figure}

\subsection{Mean Squared Error (MSE) Analysis}
\begin{figure}[htbp]
\centering
\includegraphics[scale=0.4]{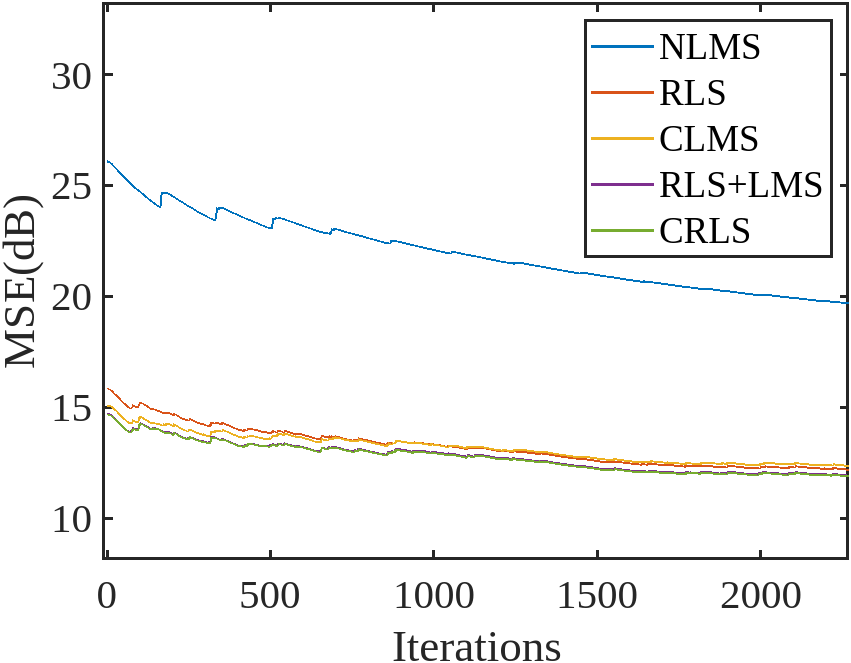}\vspace{-1mm}
\caption{Comparison of MSE \vspace{-3mm}}
\label{MSE}
\end{figure}

Figure \ref{MSE} shows the convergence of the MSE which can be calculated as 
\begin{equation}
\operatorname{MSE(n)}=\frac{1}{n} \sum_{i=1}^{n}\left(d(i)-\hat{y}(i)\right)^{2}=\frac{1}{n} \sum_{i=1}^{n}\left(e(i)\right)^{2}
\end{equation}
where $d(n)$ and $\hat{y}(n)$ are the desired and output of the filter respectively. The goal is to minimize $E((d-\hat{y})^2)$. It can be seen that any combination of adaptive filters converges to a lower value than any single Normalized Least-Mean-Square (NLMS) filter and RLS adaptive filter. The MSE curves of RLS-LMS and CRLS almost overlap but the MSE of CRLS is still lower. Overall, the MSE of CRLS is the lowest among these five adaptive filters.\vspace{-2mm}

\vspace{-6mm}\subsection{Waveform Analysis}\vspace{-1mm}
Figures \ref{fig:clms}, \ref{fig:crls}, and \ref{fig:rls_lms} show the AECG and FECG from samples 1600-2400 filtered by different combination filters. The blue points are peaks of FECG extracted by the filter. And the red circles are the MECG signal that should be suppressed. It is clear that the original high amplitude MECG components in the AECG are well eliminated in the FECG. And the amplitude of the rest component is lower than the peak of AECG which means that it can be used through the method of peak-to-peak analysis. Also, the peak of the FECG is periodic, which is consistent with the characteristics of ECG signals. As shown in Fig \ref{fig:rls} and \ref{fig:nlms}, FECG filtered by a single adaptive filter is not successful because even though the amplitude is attenuated, the MECG is not completely eliminated. As a result, the MECG signal is greater than the peak of FECG, making the FECG indistinguishable. Therefore, a combination of adaptive filters is way more effective in extracting higher-quality FECG signals than a single adaptive filter. In addition, CRLS shows the best noise suppression performance of any other combinations as shown in Fig \ref{fig:crls}.

\begin{table}
\centering
\begin{tabular}{ccc}
\hline
\textbf{Parameters(\%)} & \textbf{CRLS} & \textbf{RLS}\\
\hline
Sensitivity&100&100\\
Accuracy& 95.45&95.45\\
PPV &95.45&95.45\\
F1 score &97.67&97.67\\
\hline
\end{tabular}\vspace{-2mm}
\caption{Filter Performance on Channel 1 Signals}
\label{table:3}
\end{table}

\begin{table}
\centering
\begin{tabular}{cccc}
\hline
\textbf{Parameters(\%)} & \textbf{MSF-ANC}\cite{n} & \textbf{RLS} & \textbf{CRLS} \\
\hline
Sensitivity  &91.66  &\textbf{100} & 95.24\\
Accuracy        &84.61  &62.86 & \textbf{87}\\
PPV           &\textbf{91.66}  &62.86 &90.91\\
F1 score     &91.66  &77.19 &\textbf{93.02} \\
\hline
\end{tabular}\vspace{-2mm}
\caption{Filter Performance on Channel 2 Signals}\vspace{-4mm}
\label{table:2}
\end{table}

\subsection{Quantitative Metrics}
The quantitative metrics can be evaluated by Pan-Tompkins R-Peaks detection, which is able to recognize QSR complexes automatically~\cite{pan1985real}.  
Table \ref{table:2} and table \ref{table:3} compare the quality of the FECG extracted by CRLS and RLS when using different input channels against all metrics, along with the previous SOTA results obtained using the Multiple Sub-Filter Adaptive Noise Canceller (MSF-ANC) filtering methods which are only available on channel 2~\cite{n}. We compared channel 1 and channel 2 as input signals with reference signals coming from channel 8. 
The experimental results in Table \ref{table:3} show that CRLS and RLS attain the exact same performance with channel 1 signals, which have strong FECG signals and thus are easier to extract, demonstrating the validity of the combined filter.
Channel 2 has a small AECG and a large MECG, making the raw SNR very low and thus the extraction of FECG challenging. Table \ref{table:2} shows the Sensitivity, Accuracy, Positive Predictive Value (PPV), and F1 score of CRLS and MSF-ANC~\cite{n} on Channel 2 signals. First, CRLS significantly outperforms the single RLS filter, demonstrating a stronger denoising ability. Furthermore, compared with MSF-ANC, the sensitivity, accuracy, and F1 score of CRLS are improved by 3.58\%, 2.39\%, and 1.36\%, respectively.
This indicates that CRLS has a stronger anti-interference ability and is more advantageous for extracting low-power signals from high-power noise in extraction tasks.

\vspace{-3mm}\section{Conclusion}\vspace{-3mm}
In this paper, we introduced a new affine combination algorithm for adaptive filters to extract FECG signals from mixed AECG signals. Our algorithm identified CRLS as a highly effective filter, and we verify the feasibility of CRLS filter on the FECG signal extraction task. Compared results with the SOTA MSF-ANC method, the sensitivity, accuracy, and F1 score are improved by 3.58\%, 2.39\%, and 1.36\%, respectively. Especially in low raw SNR scenarios, CRLS has the ability to suppress high-power noise while extracting weak signals of interest, demonstrating the robustness of the CRLS filter. Our affine combination algorithm is highly efficient at extracting FECG in high-noise scenarios, providing an effective tool for not only obstetricians but also a promising next direction for the detection of weak signals in other complex environments.

\bibliographystyle{IEEEtran}
\clearpage

\begin{footnotesize}
\bibliography{reference}
\end{footnotesize}
\end{document}